\documentstyle[aps,prl,twocolumn,epsf]{revtex}
\newcommand{\p}{\mbox{$^{\prime}$}}
\newcommand{\cpp}{\mbox{$\chi^{\prime\prime}$}}
\newcommand{\lvo}{LiV$_2$O$_4$}
\begin{document}
\input{psfig.sty}
\draft
\twocolumn[\hsize\textwidth\columnwidth\hsize\csname
@twocolumnfalse\endcsname

\title{\bf
Spin fluctuations in a magnetically frustrated metal $\rm \bf LiV_2O_4$}
\author{S.-H.  Lee$^{1,2}$, Y. Qiu$^3$,
C. Broholm$^{3,2}$, Y. Ueda$^{4}$, and J. J. Rush$^{2}$}
\address{$^{1}$Department of Physics, University of Maryland, College Park, Maryland 20742}
\address{$^{2}$NIST Center for Neutron Research, National
Institute of Standards and Technology, Gaithersburg, MD 20899}
\address{$^3$Department of Physics and Astronomy,
The Johns Hopkins University, Baltimore, MD 21218}
\address{$^4$Institute for Solid State Physics, University of Tokyo,
Roppongi, Minato-Ku, Tokyo 106, Japan}
\maketitle
\begin{abstract}
Inelastic neutron scattering is used to characterize spin fluctuations
in the $d$-electron heavy fermion spinel $\rm LiV_2O_4$.  The spin relaxation rate, $\Gamma_Q$, for $Q=0.6$
\AA$^{-1}$ is 1.4(2) meV at low temperatures and increases linearly
with temperature at a rate of 0.46(8) k$_B$. There is antiferromangetic
short range order at low temperatures with a characteristic wave vector
$Q_c= 0.64(2)$ \AA$^{-1}$ and a correlation length of 6(1) \AA . While warming shifts intensity towards
lower $Q$, the staggered susceptibility peaks at a finite wave vector for $T<80$ K. The data are compared with conventional
heavy fermion systems, geometrically frustrated insulating magnets, and
recent theories for $\rm LiV_2O_4$.

\end{abstract}


]

\newpage

Owing to geometrical frustration, magnetic B-site spinel systems $\rm
AB_2O_4$ consistently display unusual magnetic properties\cite{anderson,villain,ramirev}. Ti\cite{liti2o4,hohl}, V\cite{ueda97,kondotak}, and Cr\cite{zncr2o4}
spinels are particularly interesting because all 3d electrons occupy
$t_{2g}$ orbitals that do not hybridize with oxygen orbitals\cite{good}.
Accordingly, nearest neighbor magnetic exchange interactions dominate and this maximizes frustration. In addition such materials have only 3d $t_{2g}$
bands near the chemical potential and this leads to strong correlations
in the charge sector as well. When the tetrahedral cation is
di-valent (A=Zn, Mg) the octahedral B-site has integral valency and the materials
are Mott insulators with geometrically frustrated magnetism\cite{zncr2o4}. When the
A-site is monovalent (Li) the B-site has non-integral valence resulting in a narrow-band metal. The combination of a strongly correlated metal and a spin system that cannot
order due to geometrical frustration leads to materials with complex
and unusual physical properties.

\lvo~ is a case in point\cite{kondo97}. A paramagnetic metal for $T>0.01$ K, the Sommerfeld constant increases below $T=30$ K reaching the largest value ever recorded in a d-electron system: 0.42 J mole$^{-1}$K$^{-2}$ for $T=1.5$
K\cite{Johnston99}.  The susceptibility is also large at low $T$ and the Wilson ratio
un-renormalized. To determine the origin of heavy fermion behavior in \lvo , we measured the generalized spin susceptibility as a
function of energy ($\hbar\omega$) and wave vector transfer ($Q$) using inelastic neutron
scattering. At low temperatures ($T$), the response is indistinguishable from
conventional rare earth or actinide heavy fermion system. $\chi\p\p (Q,\omega)$  peaks at a finite $Q=0.64(2)$ \AA$^{-1}$ and there is a small finite spin-relaxation rate, $\Gamma=1.4(2)$ meV. On
warming, the relaxation rate increases linearly with $T$
at a rate of $0.46(8)k_B$. Apart from the absence of a low $T$ phase transition, this behavior is reminiscent of insulating frustrated magnets. Our results strengthen the case that frustration is central to the physics of \lvo .

Since the discovery of heavy fermion behavior in \lvo\cite{kondo97},
significant experimental\cite{Johnston99,takagi,Johnston2000,krimmel99}
and theoretical\cite{anisimov99,eyert99,matsuno99,singh99,varma99,fulde} attention has been devoted to it. LDA band structure calculations\cite{anisimov99,eyert99,matsuno99,singh99} have shown that 3d $t_{2g}$
orbital triplet bands are indeed alone in crossing the fermi level.
While the trigonal splitting of the orbital triplet is less than 0.1 eV, the
lower lying $A_{1g}$ singlet has a bandwidth of only 1 eV compared to
the 2 eV bandwidth of the $E_{g}$ doublet\cite{anisimov99}. These observations have led to
suggestions that an Anderson like model might describe \lvo\cite{anisimov99,varma99} with the
half filled $A_{1g}$ singlets playing the role of localized spins and
the  quarter filled $E_{g}$ doublet acting as the conduction
band. While frustration plays a secondary role in these models, a recent paper by Fulde et al.\cite{fulde} proposes that the vanadium lattice of corner-sharing tetrahedra frustrates charge ordering and leads to isolated rings and finite length spin-1/2 and spin-1 chains. The latter have a Haldane spin gap, the former gapless excitations that are to account for the large linear term in the specific heat.  A previous neutron scattering experiment
reported a ferromagnetic (FM) to antiferromagnetic (AFM) cross over in \lvo \cite{krimmel99}. Our more complete data show that the staggered susceptibility peaks at a finite wave vector for $T<80$ K and that the heavy fermi liquid develops from a cooperative paramagnet with short range AFM spin correlations.

A 40 g powder sample of \lvo~was prepared by a solid state reaction
technique described elsewhere\cite{ueda98}. To reduce neutron
absorption from $^6$Li we used 98.5\% $^7$Li enriched starting
materials. Rietveld refinement of neutron diffraction data showed
that the sample is single phase spinel space-group $Fd\bar{3}m$ with lattice parameter
$a=8.227$\AA\ for $T<70$ K\cite{chma97} and oxygen on $32e$ sites
with $x$=0.26122(2).  We used the NIST cold neutron triple-axis spectrometer SPINS, with a large flat analyzer and a Position-Sensitive detector (PSD) to enhance the data collection rate.  Full Width at Half Maximum
energy resolution was 0.1 meV$<\Delta E<0.15$ meV and angular resolution
$\Delta 2\theta\approx 50\p$.  Background was measured for all wave vectors and energy
transfers reported. For large scattering angles ($2\theta >20^o$)
this was done by detuning the analyzer. Smaller angle backgrounds were measured with the analyzer in reflection mode but no sample in the beam.
The absolute efficiency of the
instrument was 

%==============================Fig.1==================================
\vspace{0.15in}
\noindent
\parbox[b]{3.4in}{
\psfig{file=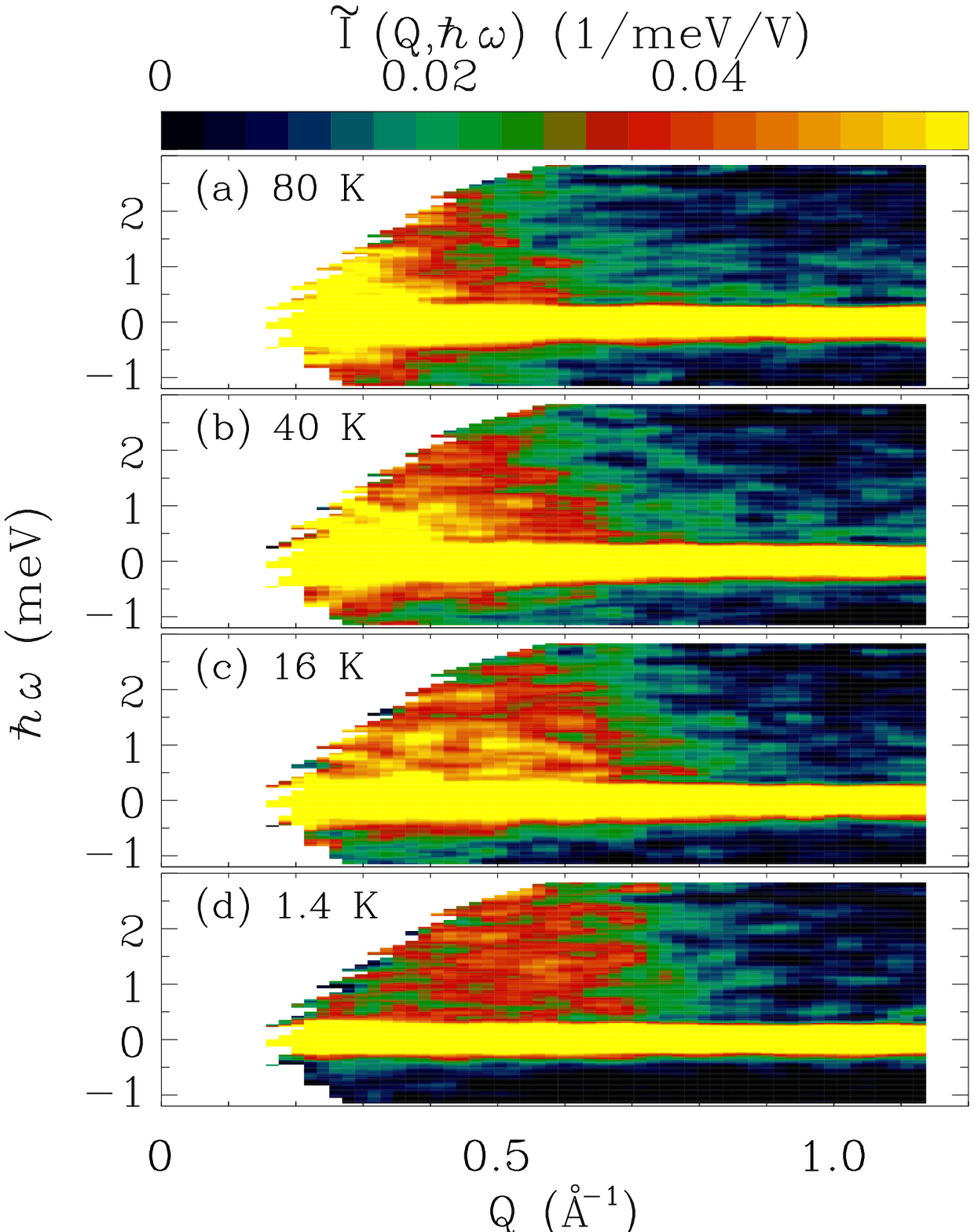,width=3.2in}
{Fig.~1. \small
Images of neutron scattering intensity from \lvo\
versus wave vector and energy transfer at four temperatures.}
}
\vspace{0.05in}
%=====================================================================

\noindent
measured to an accuracy of 15\% using the (111) nuclear Bragg peak.  The corresponding normalization
factor was applied to background subtracted data to extract the normalized magnetic neutron scattering intensity\cite{lovesey}
\begin{eqnarray}
\tilde{I}(Q,\omega )=\int\frac{d\Omega_{\hat{\bf Q}}}{4\pi}
|\frac{g}{2}F(Q)|^2 \sum_{\alpha\beta}(\delta_{\alpha\beta}-\hat{Q}_\alpha
\hat{Q}_\beta ) {\cal S}^{\alpha\beta}({\bf Q},\omega )\nonumber .
\end{eqnarray}
Here $F(Q)$ is the magnetic form factor and ${\cal
S}^{\alpha\beta}({\bf Q},\omega )$ is the dynamic spin correlation
function.  For accurate determination of low $T$
relaxation rates (Figs. 2 and 3), we derived and then subtracted the $T$ independent elastic line shape
by requiring that the resulting inelastic scattering satisfy detailed
balance at all $T$\cite{thesis}.

Fig. 1 provides an overview of the data in the form of images
of $\tilde{I} (Q,\omega)$ at four different temperatures.  The
horizontal yellow line is elastic nuclear incoherent scattering.
Magnetic scattering is apparent at finite energy transfer and for $Q< 1$ \AA $^{-1}$. In accordance with detailed
balance, warming shifts intensity from positive to negative energy
transfer.  At low temperatures magnetic scattering is strongest at
finite $Q$. At higher temperatures inelastic magnetic
scattering increases monotonically with decreasing $Q$. The total magnetic scattering cross
section is a measure of the fluctuating moment $\langle\delta m^2\rangle$ within the dynamic range
of the experiment.
Integrating the low $T$ data over $\hbar\omega~\epsilon~ [0.2,3]$ meV and 
$Q ~\epsilon~ [0.4,0.9]$ \AA$^{-1}$ we obtain 
$\langle\delta m^2\rangle=(2\mu_B)^2(3/2)\int\hbar d\omega \int
Q^2dQ  \tilde{I}(Q,\omega)/\int Q^2dQ =
0.30(2)\mu_B^2/V$. For comparison the squared effective 
%==============================Fig.2==================================
\vspace{0.15in}
\noindent
\parbox[b]{3.4in}{
\psfig{file=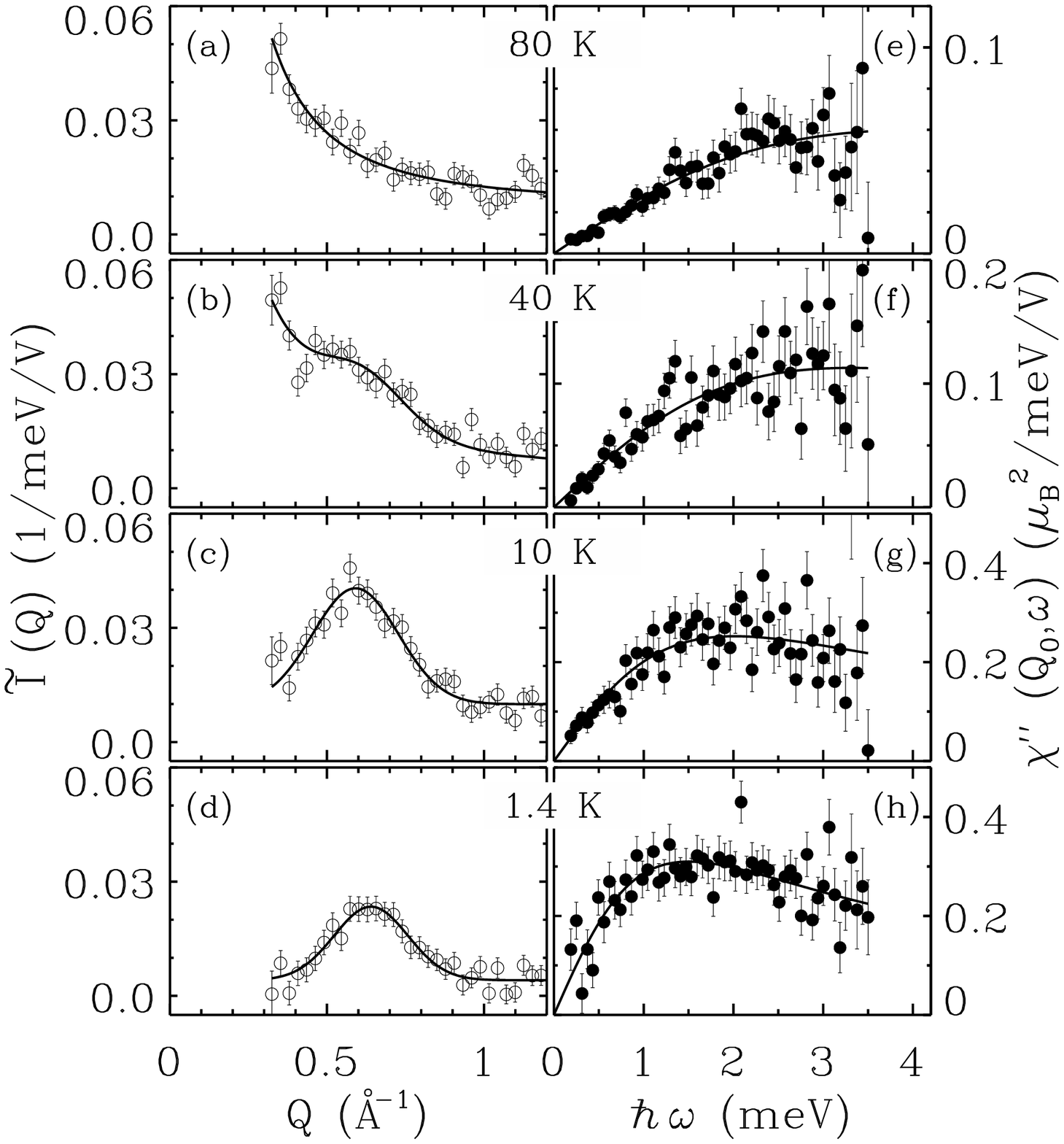,width=3.2in}
{Fig.~2. \small
(a)-(d) $\tilde{I} (Q)$ averaged over $\hbar\omega~\epsilon~[0.2,0.8]$ meV at various temperatures.  Solid lines are guides to eye. (e)-(g)
Dynamic susceptibility $\chi\p\p (Q_0,\omega)$
at $Q_0=0.6$ \AA$^{-1}$ derived from magnetic
neutron scattering data via the fluctuation dissipation theorem.
The solid lines are fits to Eq. 1. All frames show background subtracted data.
}
}
\vspace{0.05in}
%=====================================================================

\noindent
moment inferred from high temperature susceptibility measurements is 2.8$\mu_B^2$\cite{ueda97}. Our data indicate a reduced moment in the low $T$ heavy fermion phase, which is consistent with a recent high field magnetization experiment\cite{Johnston2000}. 

From the birds eye view we proceed to examine constant$-\hbar\omega$ and constant$-Q$ cuts through the data. The left column of
Fig. 2 shows the $Q$-dependence of magnetic scattering intensity
averaged over energy from 0.2 meV to 0.8 meV.
At low temperatures, there is a broad peak centered at
$Q_c= 0.64(2)$ \AA$^{-1}=0.84(3)a^*$.
Strong diffuse scattering at low $T$ is a hall-mark of
frustrated magnets. The characteristic wave vectors for this scattering are
$Q_c=0.72(2)a^*$  for $\rm Y_2Mo_2O_7$\cite{gardner},
$Q_c=1.01(2) a^*$ for $\rm ZnFe_2O_4$\cite{kamazawa},
$Q_c=1.7(1)a^*$ for $\rm ZnV_2O_4$\cite{znv2o4},
and $Q_c=1.98(6)a^*$ for $\rm ZnCr_2O_4$\cite{zncr2o4}.
For $\rm ZnFe_2O_4$ and
$\rm ZnCr_2O_4$ single crystals are available and it has been found that diffuse scattering is distributed over (different) parts
of the Brillouin zone boundary.

It is interesting to note that $Q_c$ for $\rm LiV_2O_4$
is close to values found in
$\rm Y_2Mo_2O_7$ and $\rm ZnFe_2O_4$, both materials that are expected to have longer range
interactions than {\em insulating} Cr and V spinels. A possible interpretation is that
longer range interactions are also present in {\em metallic} $\rm LiV_2O_4$.
Alternatively, and this could be checked by further analysis of existing band structure calculations, $Q_c$ for $\rm LiV_2O_4$ may result from fermi surface nesting\cite{sidis}. The peak in the Q-dependence of magnetic neutron scattering from $\rm LiV_2O_4$ has a Half Width at Half Maximum $\kappa \sim $ 0.16(2)~\AA$^{-1} \sim 0.24 a^*$ at $T=1.4$ K corresponding to an AFM correlation length of only $\xi \sim 6(1)$ \AA $=2.1(3) d_{V-V}$. Warming shifts intensity towards lower wave vectors until at $T=80$ K $\tilde{I} (Q)$
decreases monotonically with increasing $Q$ in the range probed.

The right column of Fig.~2 shows the corresponding excitation spectra.
Following background subtraction, we used the fluctuation dissipation
theorem\cite{lovesey}
to extract the imaginary part of the staggered spin susceptibility,
$\cpp (Q ,\omega )$ from the data. Though the data were acquired at the ``critical'' wave vector $Q_c=0.6$ \AA$^{-1}$, $Q-$averaged data were similar. Nonetheless, significant relaxation rate dispersion may be masked due to powder averaging.
To parametrize the $T$ dependent spectra we used the following single imaginary pole susceptibility
\begin{equation}
\chi\p\p(Q,\omega) = \frac{ \chi_{Q} \Gamma_{Q}\omega}{\omega^2+\Gamma_{Q}^2}.
\end{equation}
Here $\Gamma_{Q}$ is the relaxation rate and
$\chi_Q$ is the static staggered susceptibility.
The lines through the data in Fig.~2 (e)-(h), show that
Eq. 1 provides an adequate description of the
spectra, as it does in many conventional heavy fermion systems\cite{cb_review}.

To correlate changes in the $Q-$dependence of scattering
with anomalies in bulk
properties, Fig.~3 shows the $T-$dependence of inelastic magnetic neutron
scattering at $Q=0.6$ \AA$^{-1}$ and 0.35 \AA$^{-1}$. Evolution from a finite
$Q$ maximum to monotonically decreasing intensity with $0.35<Q<1$ \AA$^{-1}$ at $\hbar\omega=0.5$ meV occurs in the same general temperature range as the
low $T$ increase in the Sommerfeld ratio $C/T$. Thus as in conventional heavy fermion systems\cite{cb_review}, the cross over to a coherent heavy fermi liquid is associated with the development of low energy, short-range AFM correlations. The squares
show the difference between normalized horizontal and vertical field spin flip scattering
at $Q=0.35$ \AA$^{-1}$ and provide proof that the low $Q$ scattering is indeed magnetic\cite{mrk}.

Fits to the data in Fig. 2 (e)-(h) yield the $T$ dependent staggered susceptibility and relaxation rate shown in Figs. 3 (b) and (c). Comparison to bulk ($Q=0$) susceptibility
data (open symbols, frame (b)), reveals that the staggered susceptibility increases beyond the uniform susceptibility around $T=80$ K.
Thus the Curie Weiss temperature $|\Theta_{CW}|=63$ K, is the temperature scale for the appearance of short-range AFM correlations as is common for frustrated magnets\cite{gardner}.
The solid line in Fig. 2 (b) shows that $\chi_{Q_c}(T)$ can be described by a Curie-Weiss law at high temperatures:
$\chi_{Q_c}(T)=(\mu_{Q_c}^2/3k_B\theta)/(1+T/\theta )$
with $\mu_{Q_c}=1.6(1)\mu_B$, and  $\theta=7.5(10)$~K. 
The reduced Curie Weiss temperature 
%==============================Fig.3==================================
\vspace{0.05in}
\noindent
\parbox[b]{3.4in}{
\psfig{file=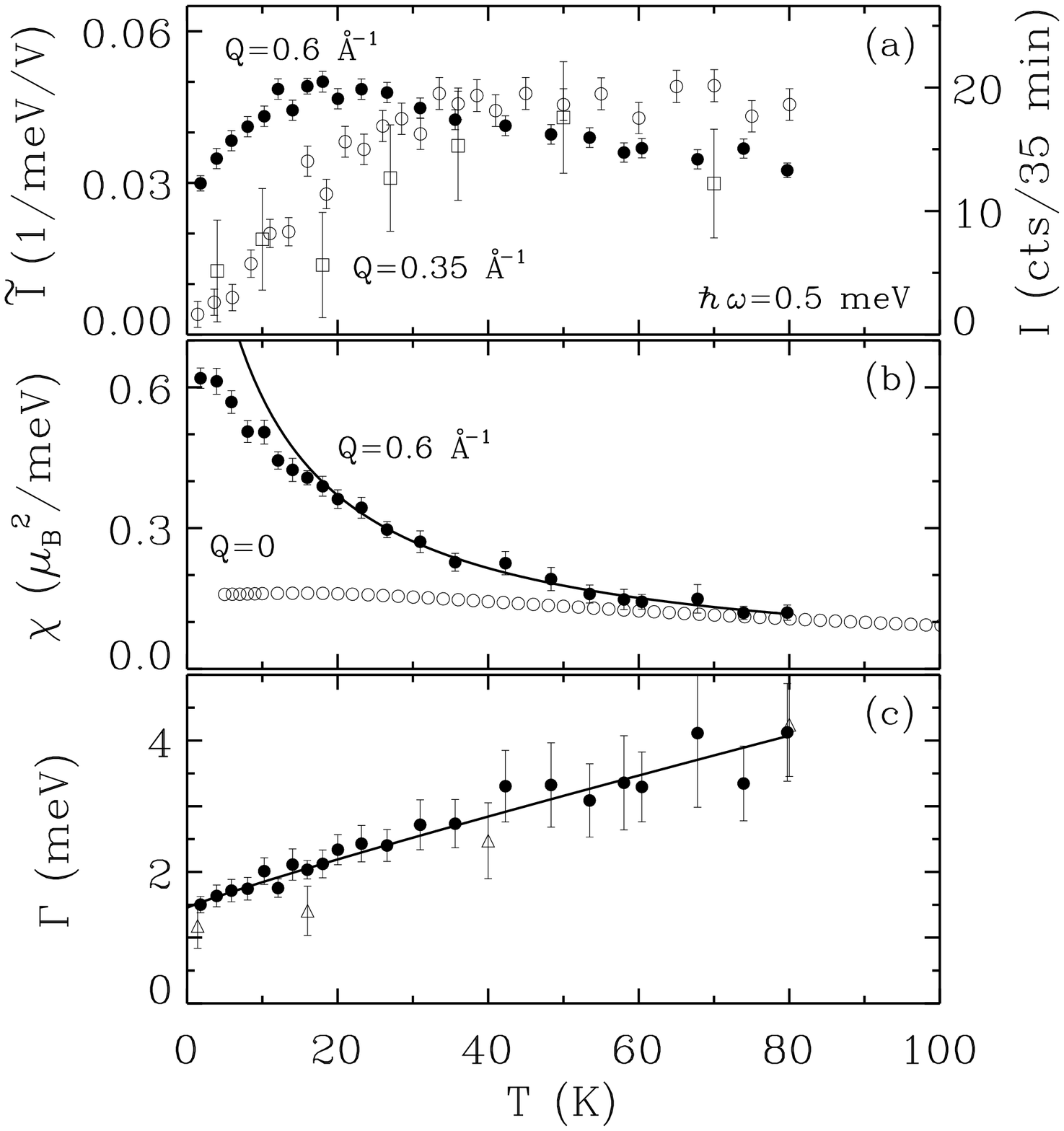,width=3.1in}
{Fig.~3. \small $T$ dependence of the following: (a)
inelastic scattering intensity at two $Q$ and $\hbar\omega=0.5$ meV. Open squares show the difference between horizontal and vertical field spin-flip neutron scattering at $\hbar\omega=0.5$ meV and  $Q=0.35$ \AA$^{-1}$. The data are normalized (left scale) and raw difference counts are on right axis. (b) bulk ($Q=0$) susceptibility and staggered susceptibility derived from fits to Eq. 1. (c) Relaxation rate from the same fits. Solid circles: $Q_c=0.6$ \AA $^{-1}$, triangles $Q-$integrated data ($0.6<Q<1.3$ \AA$^{-1}$).
}}
\vspace{0.05in}
%=====================================================================

\noindent
shows that magnetic neutron scattering at $Q_c$ is closer
to accessing critical fluctuations than bulk $Q=0$ susceptibility data.
The deviation from Curie Weiss behavior for $T<T_c\approx 20$ K (also seen at $Q=0$) indicates the development of correlations beyond nearest neighbors at the crossover to the coherent heavy fermion phase.

The relaxation rate versus temperature is shown in Fig. 3 (c). Solid and open symbols show fits to $Q=0.6$ \AA$^{-1}$ data and data integrated over the spherical shell in $Q$ space 0.6 \AA$^{-1}<Q<1.3$ \AA$^{-1}$. As previously mentioned, the ``critical'' $Q=Q_c$ and ``local'' response have similar relaxation rates. Note however that a single-crystal is required to access the actual critical wave {\em vector} where the relaxation rate could be considerably reduced compared to powder data that only specify $|Q|$.
The solid line in Fig. 3 (c) is a fit to
$\Gamma_Q (T)=\Gamma (0)+{\cal C}k_B T(T/\theta)^{\alpha-1}$.
The fit is to the higher quality $Q=0.6$ \AA$^{-1}$ data but
similar numbers describe the wave vector averaged spin relaxation rate.
We first discuss the residual $T\rightarrow 0$ relaxation rate that
refines to $\Gamma_Q (0)=1.4(2)$ meV. In a crude band description of
heavy fermion behavior, the product of the Sommerfeld constant and
the low $T$ spin relaxation rate, $\Gamma (T=0)$,
should be proportional to the electron density but independent
of the bandwidth. The low $T$ limit of this product
is 0.60(4) meV J mole$^{-1}$K$^{-2}$=0.85(6) k$_B^2$/f.u. for $\rm LiV_2O_4$ compared to an average of 1.0 $k_B^2$/f.u. for cerium heavy fermion systems\cite{regnault88}. The numbers are consistent with a common link between spin fluctuations and the enhanced specific heat of these materials.

The $T$ dependence of
$\Gamma_Q(T)$ in $\rm LiV_2O_4$ is close to linear  (${\cal C}=0.46(8)$
and $\alpha=0.9(2)$) in the range probed. While linear $T$
dependence of the spin relaxation rate is not unheard of in these materials, actinide and rare
earth heavy fermion systems often display sub-linear $T$
dependence for $T\approx \Gamma (T=0)/k_B$\cite{regnault88}. This
behavior is generally associated with an unstable local moment.
Linear $T$ dependence of the spin relaxation rate is however common among frustrated transition metal oxides.
$\rm LiV_2O_4$ has the largest Sommerfeld constant of the
transition metal oxides. It is the only known vanadium oxide to
remain cubic at low temperatures, and the compound borders an
entropic spin glass phase in $\rm Li_xZn_{1-x}V_2O_4$ for
$0.1<x<0.9$\cite{urano}. Our data for $\Gamma_Q (T)$ provide an additional indication that the geometrically frustrated physics of spins on a lattice of corner-sharing tetrahedra is the crux of heavy fermion behavior in $\rm LiV_2O_4$.

Theories of heavy fermion systems are generally non-lattice specific\cite{hfreview}.
While Fulde et al.\cite{fulde} were the first to focus on frustration,
the theory appears to be inconsistent with our scattering data.
The characteristic wave vector for fluctuations of spin-1/2 chain-lets
and rings would be $\pi/d_{V-V}=\sqrt{2}a^*$, which is quite different
from $Q_c=0.84(3)a^*$ observed in Fig. 2 (d). Previous neutron scattering data from \lvo\ were interpreted as indicating a cross over from AFM to FM fluctuations on heating above $T=40$ K\cite{krimmel99}. Fig. 3(b) shows that the uniform susceptibility is less than the staggered susceptibility for $T<80$ K, a clear indication that \lvo\ enters the heavy fermion phase as an AFM correlated paramagnet. The apparent FM response in $\rm LiV_2O_4$ may be a consequence of powder averaging scattering from a spin system where $\kappa>Q_c$.
The previous scattering paper also concluded that the fluctuation rate of low Q decreased dramatically with increasing $T$. The present results indicate that this conclusion was an artifact of analyzing spectra taken at constant scattering angle\cite{murani}. Constant-$Q$ spectra decouple changes in spatial and temporal correlations and show that the relaxation rate increases continuously in proportion to $T$ with no significant anomaly associated with entering the fermi liquid phase.

In summary, our experiment has revealed features both of a strongly correlated metal and of frustrated magnetism in \lvo .  Short range correlations for $T<<|\Theta_{CW}|$ and a linear rise in spin relaxation rate with $T$ are common features of insulating frustrated magnets. A residual low $T$ relaxation rate, a reduced effective moment at low energies, and an increasing AFM correlation length upon entering the coherent fermi liquid phase are features of a strongly correlated metal. Further progress towards understanding the mix of magnetic frustration and correlated electrons in \lvo\ will require neutron scattering experiments on single crystalline samples and a lattice specific theory of heavy fermions.

We thank Y.B. Kim and D. Huse for helpful discussions. Work at SPINS is based upon activities supported by the NSF under
DMR-9986442. Work at JHU was supported by the NSF through DMR-0074571.

\end{document}